\documentstyle[preprint,aps,epsfig]{revtex} \begin{document}
\title{Pairing Instability and Mechanical Collapse of a Bose Gas with an
Attractive Interaction} \author{Gun Sang Jeon$^{1,2}$, Lan Yin$^1$, Sung
Wu Rhee$^1$, David J. Thouless$^1$} \address{$^1$Department of Physics,
University of Washington, Seattle, WA 98195-1560} \address{$^2$Center for
Strongly Correlated Materials Research, Seoul National University,\\ Seoul
151-742, Korea} \date{\today} \maketitle \begin{abstract} We study the
pairing instability and mechanical collapse of a dilute homogeneous bose gas
with an attractive interaction.  The pairing phase is found to be a saddle
point and unstable against pairing fluctuations.  This pairing saddle point
exists above a critical temperature.  Below this critical temperature, the
system is totally unstable in the pairing channel.  Thus the system could
collapse in the pairing channel in addition to mechanical collapse.  The
critical temperatures of pairing instability and mechanical collapse are
higher than the BEC temperature of an ideal bose gas with the same density.
When fluctuations are taken into account, we find that the critical temperature
of mechanical collapse is even higher.  The difference between the collapse
temperature and the BEC temperature is proportional to $(n|a_s|^3)^{2/9}$,
where $n$ is the density and $a_s$ is the scattering length. \end{abstract}

\newpage The system of Bose
gas with attractive interaction has received a lot of attention recently due
to the success in cooling $^7$Li\cite{Li} and $^{85}$Rb\cite{Rb} systems.  In
these trapped systems, Bose Einstein condensation can be realized at zero
temperature if the number of atoms in the condensate is below a critical
value\cite{trap}, because of a finite size effect.  In a $^7$Li system,
the condensate experiences repetitive growth and collapse as the system is
cooled down\cite{Li}.  In a $^{85}$Rb system, the interaction can be
varied from being repulsive to being attractive by tuning a Feshbach
resonance.  During this process, the system loses particles until the
critical number is reached\cite{Rb}.

A homogeneous Bose gas with attractive interaction is generally unstable
at zero temperature because the system can always lower energy by
increasing density.  At high temperatures entropy stabilizes the gaseous
phase.  There are several scenarios as to how the system becomes unstable
when the temperature is lowered.  The simplest picture is that the system
collapses within the normal gaseous phase, which is supported by several
theoretical studies\cite{collapse}.  This collapse is mechanical collapse,
where the compressibility becomes negative and the system can lower energy
by separating into a low density phase and a high density phase. Another
possibility is that the system goes into the Bose-condensed phase and then
collapses in the condensed phase.  However the condensation transition
temperature is found to be lower than that of mechanical
collapse\cite{collapse}.  In addition to these scenarios, Evans and Imry
proposed that there is a pairing phase similar to the BCS phase in
superconductors at low temperature.  However it was also concluded in the
previous theoretical studies\cite{collapse} that this pairing phase
transition temperature is lower than the mechanical collapse temperature
of the normal phase.

It was not fully addressed in the previous studies whether the
pairing phase could exist as a metastable phase or not.  In this paper, we
examine its stability in detail.  In addition, we go beyond the Hartree-Fock
approximation to see how the temperature of mechanical collapse
is affected by fluctuations.  

The interaction between atoms is singular at short distances.  However, as
for other cases in condensed matter and elementary particle physics, at
low energies the dilute Bose gas system can be well described by the
Hamiltonian with an effective contact interaction\cite{pp} with the
coupling constant proportional to the scattering length $a_s$, up to a
momentum transfer which is determined by the effective range of the
interaction.  The unphysical ultraviolet divergences in this approach can
be subtracted by the pseudopotential method\cite{pp}.  This
renormalization scheme is justified in the case of a dilute Bose gas with
weak interaction at zero temperature, where it can be shown that the
divergences come from double counting of certain diagrams and the dominant
perturbations come from the repetitive scatterings of two particles.

At finite temperature when there is no condensate, the medium effect
becomes important and the coupling constant could depend on other quantities
in addition to $a_s$.  To avoid this complexity, here we use a simple model
with a contact interaction which has coupling constant $g$ and an explicit
cutoff $\Lambda$ in the wave-vector space.  The Hamiltonian of the system is
given by
\begin{equation}
H={\hbar^2 \over 2m}\nabla\psi^\dagger\cdot\nabla\psi+g{\psi^\dagger}^2 \psi^2.
\end{equation}
A length scale $a$ can be defined from the coupling constant,
$a\equiv m/(4\pi\hbar^2g)$, but $a$ is not the scattering length $a_s$.
By solving the two-body scattering problem, we obtain the relation between
$a$ and $a_s$, $a_s=a/(1+2a\Lambda/\pi)$.  The cutoff $\Lambda$ is
proportional to $1/\sqrt{|a_s|r_s}$, where $r_s$ is the effective range.  Here
we consider the case of a weakly attractive interaction with $a<0$,
$-2a\Lambda/\pi<1$, and $|na^3| \ll 1$, where $n$ is the density.

To study this system, we use the Peierls variational method\cite{book} which
gives an upper bound for the free energy or grand thermodynamic potential of a
quantum system.  To discuss both the normal phase and the pairing phase, we
assume the quasiparticles are superpositions of particles and holes,
and they have certain occupation numbers at finite temperature.  In momentum
space, the density matrix is given by $<\psi_{\bf k}^{\dagger}
\psi_{\bf k}>=(f_k+1/2) \cosh \theta_k-1/2$, and $<\psi_{\bf k} \psi_{\bf -k}>
=(f_k+1/2)\sinh\theta_k$, where $\cosh \theta_k$ and $\sinh\theta_k$ are
coherence factors and are chosen to be real.  The occupation number is
given by $f_k$.

Within this approach, the expectation value of the grand potential is given by
\begin{eqnarray*}
\Omega(\{\theta_k\},\{f_k\})=\sum_{\bf k}(\epsilon_k-\mu)[(f_k+1/2)\cosh
\theta_k-1/2] +{g\over 2V} \{\sum_{\bf k} (f_k+1/2)\sinh \theta_k\}^2\\
+{g \over V}\{\sum_{\bf k}[(f_k+1/2)\cosh \theta_k-1/2]\}^2
-T\sum_{\bf k}[(f_k+1)\ln(f_k+1) -f_k\ln f_k].
\end{eqnarray*}
The normal phase and the pairing phase are given by the saddle points of the
variational space,
$\partial \Omega / \partial f_k=0$,
$\partial \Omega / \partial \theta_k=0$.
The saddle point equations can be further written as
\begin{eqnarray} \label{parameter1}
\tanh\theta_k &=& {\Delta \over \epsilon_k-A-\mu},\\
\label{parameter2}
f_k &=& {1 \over 2} \coth {E_k \over 2T}-{1 \over 2},
\end{eqnarray}
where
$$\Delta=-{g \over V} \sum_{\bf k}(f_k+1/2)\sinh\theta_k,$$
$$A=-{2g \over V} \sum_{\bf k} [(f_k+1/2)\cosh \theta_k-1/2],$$
$E_k=\sqrt{\eta_k^2-\Delta^2}$, and $\eta_k=\epsilon_k-A-\mu$.
The pairing phase corresponds to the solution with finite $\Delta$
and the normal phase corresponds to the solution with $\Delta=0$.
The transition from normal phase to pairing phase occurs when
$\Delta$ approaches zero.

To simplify the discussion without compromising the saddle point
structure, we look at a more restricted variational space defined by
Eq.(\ref{parameter1}) and Eq.(\ref{parameter2}) with $\Delta$
and $A$ as variational parameters.  In terms of
these new variational parameters, the grand potential is given by
\begin{equation}
\Omega(\Delta,A)=\sum_{\bf k}(\epsilon_k-\mu)\Bigl[ {\eta_k \over 2E_k}
\coth {E_k \over 2T} -{1 \over 2}\Bigr] +{V \tilde\Delta^2 \over 2g}
+{V \tilde A^2 \over 4g}
-\sum_{\bf k} {E_k \over 2}\coth {E_k \over 2T}+
T\sum_{\bf k}\ln\bigl(2\sinh {E_k \over 2T}\bigr),
\end{equation}
where the parameters $\tilde\Delta,\tilde A$ are defined by
$$\tilde\Delta(\Delta,A) =-{g \over V} \sum_{\bf k}(f_k+{1 \over 2})
\sinh\theta_k=-{g \over V} \sum_{\bf k} {\Delta \over 2E_k}
\coth {E_k \over 2T},$$
and
$$\tilde A(\Delta,A)=-2{g \over V}\sum_{\bf k} \bigl[(f_k+{1 \over 2})\cosh
\theta_k-{1 \over 2}\bigr] =-{g \over V}\sum_{\bf k}\bigl[
{\eta_k \over E_k}\coth {E_k \over 2T} -1\bigr].$$
The new saddle point equations are given by
$\tilde A=A$, $\tilde \Delta=\Delta$.

\begin{figure}
\centerline{\epsfig{file=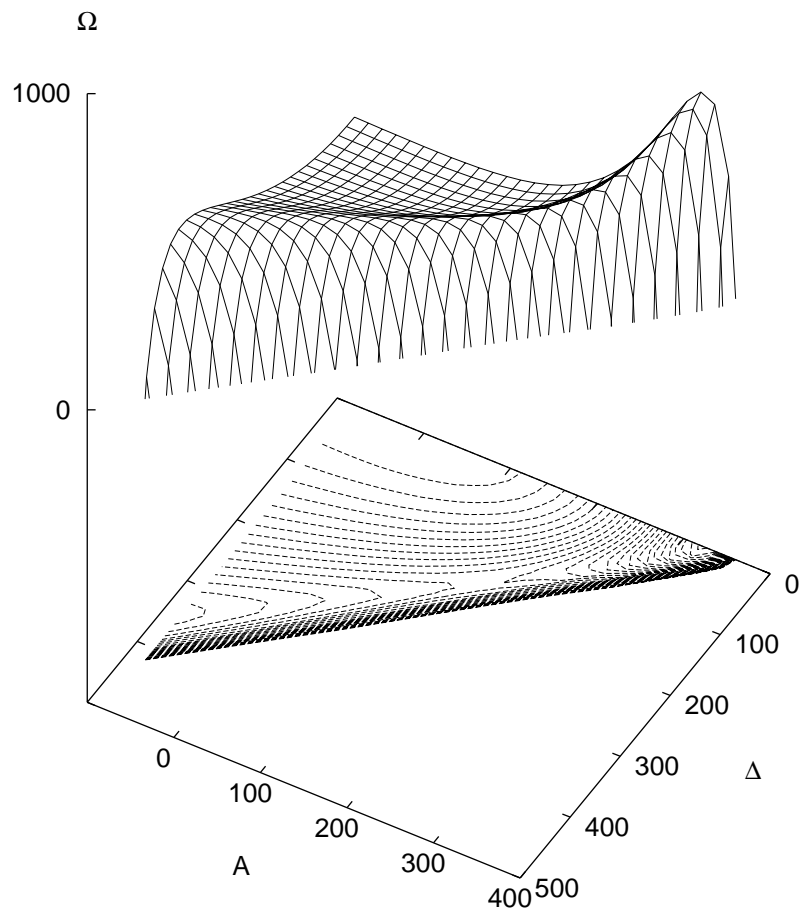}}
\caption{Grand potential for $\mu=-400$, $T=100$, and
$\Lambda=10$. 
Here the length and the energy are written in units of $|a|$
and $\hbar^2/2m|a|^2$, respectively.}
\label{Fig}
\end{figure}

The stable phases correspond to the local minimum points of the grand
potential.  If the pairing phase is stable, one of the necessary conditions 
is that it should be stable against fluctuations in pairing, which is
given by 
\begin{equation} \label{stable}
{\partial^2 \Omega \over \partial \Delta^2}
\Big|_{\tilde A=A,\tilde \Delta=\Delta}>0.
\end{equation}
Since the transition into the pairing phase would be a second-order phase
transition, at the critical temperature $T_P$,
\begin{equation}
{\partial^2 \Omega \over \partial \Delta^2}
\Big|_{\tilde A=A,\tilde \Delta=\Delta}=0,
\end{equation}
and the stability condition Eq.(\ref{stable}) becomes
\begin{equation}
{\partial^4 \Omega \over \partial \Delta^4}\Big|_{\tilde A=A,\Delta=0}>0.
\end{equation}
However, we find that the opposite inequality holds
\begin{equation}\label{IQ}
{\partial^4 \Omega \over \partial \Delta^4}\Big|_{\tilde A=A,\Delta=0}=
{3V \over 2g}\Bigl[{\partial^2 \tilde A \over \partial \Delta^2}\Bigr]^2-
\sum_{\bf k}{3 \over 2 \eta_k^3}(1+2f_k^2)<0.
\end{equation}
It clearly shows that the pairing phase is generally unstable.
The instability of the pairing phase can also be illustrated by plotting the
profile of the grand potential as shown in Fig.(\ref{Fig}).  The pairing phase
is located at a saddle point and the normal phase is located at a minimum point.
The pairing saddle point is a maximum in the direction of the normal
phase and a minimum in the perpendicular direction.

This instability can be studied in more detail by constructing an effective
Lagrangian near the critical temperature $T_P$ based on Landau's theory
of phase transitions
\begin{equation}
{\cal L}(\Delta) \equiv \Omega(\Delta,\tilde A)-\Omega(0,\tilde A) \approx
\alpha \Delta^2+\beta \Delta^4,
\end{equation}
where $\alpha={\partial^2\Omega\over2\partial\Delta^2}|_{\tilde A=A,\Delta=0}$
and $\beta={\partial^4\Omega\over24\partial\Delta^4}|_{\tilde A=A,\Delta=0}$,
$\beta<0$ from Eq.(\ref{IQ}).  When $T>T_P$, $\alpha>0$, there is a pairing
saddle-point solution $\Delta^2=-\alpha/(2\beta)$.  When $T<T_P$, $\alpha<0$,
the normal phase becomes unstable in the pairing channel and there is no
pairing saddle-point solution.  It is clear that the system is unstable in
the pairing channel when the temperature is below $T_P$.  Even above $T_P$,
the normal phase can become unstable in the pairing channel by tunneling into
the region beyond the pairing saddle point if any place in that region has
lower energy than the normal phase.  The pairing saddle point serves as an
energy barrier in this case.  However, in our limited numerical calculations,
we have not been able to determine the condition for the normal phase to
become metastable.  When the temperature is far above $T_P$, the pairing saddle
point disappears.  In this case the effective Lagrangian description breaks
down and a better method is needed to find the general expression of the
critical temperature below which the pairing saddle point appears.

The critical temperature of pairing instability $T_P$ is given by the
following equation
\begin{equation}\label{T_P}
1=-g \int {d^3k \over (2\pi)^3} {1 \over 2 \eta_k} \coth{\eta_k \over 2T_P},
\end{equation}
where $\eta_k=\epsilon_k+2gn-\mu$.  This equation has solution only
when the temperature is close to the ideal BEC temperature $T_0$.
In this case it can be further simplified,
\begin{equation}
1+{2 a \Lambda \over \pi} \approx -g \int {d^3k \over (2 \pi)^3}
{T_P \over  \eta_k^2}
\end{equation}
and
\begin{equation}
1+{2 a \Lambda \over \pi} \approx -{T_P a\over \hbar} \sqrt{2m \over \eta_0}.
\end{equation}
From the density equation, we obtain
\begin{equation}
\eta_0 \approx {9 \over 16\pi} \zeta^2({3 \over 2}) T_0 ({T \over T_0}-1)^2,
\end{equation}
where $$T_0={2\pi\hbar^2 \over m}({n \over \zeta({3 \over 2})})^{2 \over 3}$$
and $\zeta(x)$ is the Riemann-Zeta function.
Therefore the solution of Eq.(\ref{T_P}) is given by
\begin{equation}\label{pairing}
{T_P \over T_0}-1 \approx {2 \pi \over 3
\zeta^{4 \over 3}({3 \over 2})(1+{2 a \Lambda \over \pi})}
n^{1 \over 3} |a|={8 \pi \over 3 \zeta^{4 \over 3}({3 \over 2})}
n^{1 \over 3} |a_s|.
\end{equation}

In addition to the pairing instability, mechanical collapse can also
occur in the normal phase as discussed in several references\cite{collapse}.
It is important to find out which instability takes place first.  The
mechanical collapse happens when the compressibility goes to zero
$\partial \mu / \partial n=0$, which can be more explicitly written as
\begin{equation}\label{T_C}
-{g \over 2T_C} \int {d^3 k \over (2\pi)^3} {\rm csch}^2
{\eta_k \over 2T_C}=1,
\end{equation}
where $T_C$ is the temperature of mechanical collapse.  Similar to
Eq.(\ref{T_P}), Eq.(\ref{T_C}) can only be satisfied when $T_C$ is close
to $T_0$, and can be simplified as
\begin{equation}
1 \approx -g \int {d^3k \over (2\pi)^3} {2T_C \over \eta_k^2}
\end{equation}
and
\begin{equation}
1 \approx -{2T_C a\over \hbar} \sqrt{2m \over \eta_0}.
\end{equation}
The solution is given by
\begin{equation}\label{mc}
{T_C \over T_0}-1 \approx {16 \pi \over 3 \zeta^{4 \over 3}({3 \over 2})}
n^{1 \over 3} |a|.
\end{equation}
Similar results have been obtained in the previous studies\cite{collapse}.

By comparing the two critical temperatures Eq.(\ref{pairing}) and
Eq.(\ref{mc}), we find that when $-4a \Lambda/\pi>1$,
$T_P>T_C$; when $-4a \Lambda/\pi<1$, $T_P<T_C$.
However, this result is obtained within the variational approach, which is
equivalent to the Hartree-Fock approximation.  To obtain a better quantitative
result, fluctuations need to be considered.

There are infrared divergences in all orders of perturbation at $T_0$. To
obtain the correct renormalization near $T_0$, summing the leading divergent
terms at each order is necessary, which can be effectively performed in a
self-consistent approach.  Here we consider the simplest one-loop
renormalization with self consistency.  In the Hartree-Fock approximation, the
self-energy in the normal phase is simply given by $2gn$.  When fluctuations
are taken into account, the coupling constant $g$ is renormalized.  In one-loop
order, $g$ is re normalized by particle-particle scattering.  It is well known
that at zero temperature, such scatterings renormalize $g$ to
$4\pi\hbar^2a_s/m$.  In this approach, the renormalized coupling constant is
given by the $t$-matrix
\begin{equation}\label{tx}
t({\bf q},\Omega)=g-T\sum_{\omega}\int{d^3k \over (2\pi)^3}g
t({\bf q},\Omega) G({\bf k},\omega) G({\bf q}-{\bf k},\Omega-\omega),
\end{equation}
where $G({\bf k},\omega)=1/(i\omega-\Sigma({\bf k},\omega)-\epsilon_k+\mu)$ is
the finite-temperature Green's function.  Under this approximation, the new
self-energy is given by
\begin{equation}\label{sx}
\Sigma({\bf k},\omega)=2 T\sum_{\Omega}\int{d^3q \over (2\pi)^3}
t({\bf q},\Omega) G({\bf k}-{\bf q},\omega-\Omega).
\end{equation}

The infrared behavior of the integrals in Eq.(\ref{tx}) and Eq.(\ref{sx}) is
dominated by $\Sigma(0,0)-\mu$, which is small and finite at the collapse
temperature.  The $k$-dependence and $\omega$-dependence of $\Sigma$ just
provides a renormalization factor to the propagator in the leading order.  So
here we only consider only the constant part of self-energy,
$$\Sigma \equiv \Sigma(0,0) \approx 2 t(0,0) n,$$ and get
\begin{equation}
{g \over t(0,0)} \approx 1+g \int {d^3k \over (2\pi)^3} {1 \over 2 \eta_k'}
\coth{\eta_k' \over 2T},
\end{equation}
where $\eta_k'=\epsilon_k+\Sigma-\mu$.  Using the same technique in solving
Eq.(\ref{T_P}), we obtain $t(0,0)$ in terms of $\Sigma$
\begin{equation}
t(0,0) \approx {g \over 1+{2 a \Lambda \over \pi}+{T a\over \hbar}
\sqrt{2m \over (\Sigma-\mu)}}.
\end{equation}
Therefore the self-energy $\Sigma$ is given by the self-consistent equation
\begin{equation}\label{sigma}
\Sigma \approx {2gn \over 1+{2 a \Lambda \over \pi}+{T a\over \hbar}
\sqrt{2m \over (\Sigma-\mu)}}.
\end{equation}

The condition of mechanical collapse $\partial \mu/\partial n=0$ now leads
to the following equation
\begin{eqnarray}
1&&=-\int {d^3k \over (2\pi)^3}{1 \over 4T}{\partial \Sigma \over \partial n}
{\rm csch}^2 {\epsilon_k+\Sigma-\mu \over 2T}\\
&&\approx  -{T m\over 4\pi\hbar^3}{\partial \Sigma \over \partial n}
\sqrt{2m \over \Sigma-\mu},\label{mcc}
\end{eqnarray}
where from Eq.(\ref{sigma}) we obtain
\begin{equation}
{\partial \Sigma \over \partial n} \approx {2g [1+{2 a \Lambda \over \pi}
+{T a\over \hbar} \sqrt{2m \over (\Sigma-\mu)}] \over
[1+{2 a \Lambda \over \pi}+{T a\over \hbar} \sqrt{2m \over (\Sigma-\mu)}]^2-
{T a\over  \hbar} gn\sqrt{2m \over (\Sigma-\mu)^3}}.
\end{equation}
The solution of Eq.(\ref{mcc}) is approximately given by
\begin{equation} \label{eq1}
(1+{2 a \Lambda \over \pi})^2 \approx {8\pi\hbar a^2 nT\over
\sqrt{2m(\Sigma-\mu)^3}}.
\end{equation}
Close to $T_0$, the density equation yields
\begin{equation} \label{eq2}
\Sigma-\mu \approx {9 \zeta^2({3 \over 2}) \delta T^2 \over 16 \pi T_0},
\end{equation}
where $\delta T=T-T_0$.  Combining Eq.(\ref{eq1}) with Eq.(\ref{eq2}), we
obtain the new critical temperature of mechanical collapse $T_C'$
\begin{equation}
{T_C' \over T_0}-1 \approx 4{(2\pi|a_s|n^{1 \over 3})^{2 \over 3} \over
3 \zeta^{8 \over 9}({3 \over 2})}.
\end{equation}

The difference between $T_C'$ and $T_0$ is now proportional to
$(n|a_s|^3)^{2/9}$.  In the weakly-interacting limit, it is much bigger than
$n^{1/3}|a_s|$ which is proportional to the temperature difference in
Hartree-Fock approximation.  The strong renormalization of the collapse
temperature is a result of the enhancement of the Hartree-Fock attractive
interaction due to fluctuations.  In contrast, the critical temperature of
pairing instability is not renormalized by the $t$-matrix.  We have
used the one-loop and constant self-energy approximations which may limit
the accuracy of the numerical coefficient. Future works with better
approximations may improve the numerical prefactor and provide an estimate
of the next order term.

Our result shows that the collapse temperature is higher than the temperature
of pairing instability.  However the pairing saddle point exists above the
temperature of pairing instability.  The saddle point may still be
important to collapse dynamics if it appears above the collapse temperature.
In this case, if the normal phase is only a local minimum and not a global
minimum of the grand potential, there is a finite probability for the
system to pass the pairing saddle point and tunnel into the unstable region.
Discussing the dynamical process of collapse is beyond the scope of this paper.
However, within this variational framework, the collapse process can probably
be studied by the general method of analyzing the stability of metastable
states\cite{Langer} in the future.

We would like to thank the Aspen Center for Physics where some parts
of this work were produced.  This work is supported by the National Science
Foundation under Grant No. DMR 98-15932.

\begin{references}
\bibitem{Li}
J.M. Gerton {\it et al.}, Nature (London) {\bf 408}, 692 (2000).
\bibitem{Rb}
J.L. Roberts {\it et al.}, Phys. Rev. Lett. {\bf 86}, 4211(2001).

\bibitem{trap}
P.A. Ruprecht {\it et al.}, Phys. Rev. A {\bf 51}, 4704 (1995).

\bibitem{pairing}
J.G. Valatin and D. Butler, Nuovo Cimento {\bf 10}, 37 (1958);
W.A.B. Evans and Y. Imry, Nuovo Cimento \textbf{63}b, 155 (1969).

\bibitem{collapse}
H.T.C. Stoof, Phys. Rev. A {\bf 49}, 3824 (1994);
E.J. Mueller and G. Baym, Phys. Rev. A {\bf 62}, 053605 (2000).

\bibitem{pp}
T.D. Lee, K. Huang, and C.N. Yang, Phys. Rev. {\bf 106}, 1135 (1957).

\bibitem{book}
For a detailed introduction of this method, see D.J. Thouless,
{\it The Quantum Mechanics of Many-Body Systems} (Academic Press,
New York, 1961).
\bibitem{Langer}
J.S. Langer, Ann. Phys. {\bf 54}, 258 (1969).
\end{references}
\end{document}